\documentclass{article}
\usepackage{spconf}

\usepackage[english]{babel}

\usepackage{ifpdf}

\usepackage{cite} 
\usepackage{url}
\usepackage{hyperref}

\usepackage[pdftex]{graphicx}
\graphicspath{{./figures/}}
\usepackage{color}
\usepackage{pgf, tikz, pgfplots}
\usetikzlibrary{shapes, arrows, automata, plotmarks}
\usetikzlibrary{calc,hobby,decorations}

\usepackage[cmex10]{amsmath}
\usepackage{amsfonts, amssymb, amsthm}
\usepackage{mathrsfs}

\usepackage{algorithm,algpseudocode}
\algnewcommand{\LeftComment}[1]{\Statex \(\triangleright\) #1}

\usepackage[margin=15mm]{geometry}
\usepackage{enumerate}
\usepackage{multirow}
\usepackage{rotating}
\usepackage{subcaption}
\input{auxFiles/pennColors.sty}
\usepackage{needspace}

\input{auxFiles/mySymbol.sty}

\def\myfactor{1.0}
\def\unit{\myfactor cm}

\pgfplotsset{ ylabel near ticks,                 
              xlabel near ticks,                 
              tick label style = {font=\footnotesize},   
              label style = {font=\footnotesize}, 
              title style = {font=\footnotesize},
            }

\tikzstyle{empty node} = [ circle, 
                     draw = black,
                     text = black, 
                     minimum size = 0.8*\unit]

\tikzstyle{node} = [ empty node, 
                     fill = blue!50,
                     draw = blue!50,
                     text = white]

\tikzstyle{blue node} = [ empty node, 
                         fill = blue!50,
                         draw = blue!50,
                         text = white]

\tikzstyle{red node} = [ empty node, 
                         fill = red!50,
                         draw = red!50,
                         text = white]

\tikzstyle{green node} = [ empty node, 
                         fill = mygreen!50,
                         draw = mygreen!50,
                         text = white]

\tikzstyle{black node} = [ empty node, 
                           fill = black!40,
                           draw = black!40,
                           text = white]

\tikzstyle{empty dot} = [ circle, 
                           draw = black,
                           inner sep = 0pt,
                           anchor = center,
                           minimum size = 0.4*\unit]

\tikzstyle{dot} = [ empty dot, 
                    fill = blue!50,
                    draw = blue!50]

\tikzstyle{blue dot} = [ empty dot, 
                         fill = blue!50,
                         draw = blue!50]

\tikzstyle{red dot} = [ empty dot, 
                        fill = red!50,
                        draw = red!50]

\tikzstyle{black dot} = [ empty dot, 
                          fill = black!50,
                          draw = black!50]

\tikzstyle{edge}                 = [shorten >=1pt, shorten <=1pt]
\tikzstyle{directed edge}        = [edge, -stealth]
\tikzstyle{double directed edge} = [edge, stealth-stealth]
\tikzstyle{tight edge}                 = [shorten >=0pt, shorten <=0pt]

\tikzstyle{block} = [ rectangle,
                      minimum width = \unit,
                      minimum height = \unit,
                      fill = blue!15,
                      draw = black,
                      text = black]

\newtheorem{remark}{\hspace{0pt}\bf Remark}

\title{NONLINEAR STATE-SPACE GENERALIZATIONS OF GRAPH CONVOLUTIONAL NEURAL NETWORKS}
\name{Luana Ruiz$^{*}$, Fernando Gama$^\dagger$, Alejandro Ribeiro$^{*}$ and Elvin Isufi$^\ddagger$\thanks{Supported by USA NSF CCF 1717120, ARL DCIST CRA W911NF-17-2-0181. 
    }
}
\address{$^{*}$Department of Electrical and Systems Engineering, University of Pennsylvania, Philadelphia, USA \\
    $^{\dagger}$Electrical Engineering and Computer Sciences Department, University of California, Berkeley, USA \\
    $^{\ddagger}$Department of Intelligent Systems, Delft University of Technology, Netherlands
}
\begin{document}
\ninept
\maketitle
\begin{abstract}
Graph convolutional neural networks (GCNNs) learn compositional representations from network data by nesting linear graph convolutions into nonlinearities. In this work, we approach GCNNs from a state-space perspective revealing that the graph convolutional module is a minimalistic linear state-space model, in which the state update matrix is the graph shift operator. We show that this state update may be problematic because it is nonparametric, and depending on the graph spectrum it may explode or vanish. Therefore, the GCNN has to trade its degrees of freedom between extracting features from data and handling these instabilities. To improve such trade-off, we propose a novel family of nodal aggregation rules that aggregate node features within a layer in a nonlinear state-space parametric fashion allowing for a better trade-off. We develop two architectures within this family inspired by the recurrence with and without nodal gating mechanisms. The proposed solutions generalize the GCNN and provide an additional handle to control the state update and learn from the data. Numerical results on source localization and authorship attribution show the superiority of the nonlinear state-space generalization models over the baseline GCNN.
\end{abstract}
%


\section{Introduction}
\label{sec:intro}



Graph convolutional neural networks (GCNNs) learn a parametric map from high-dimensional data whose dependencies can be represented by a graph, e.g., biological data, financial data, and social network data \cite{newman2010networks, bullmore2009complex, jackson2010social}. The GCNN map is a compositional layered function of simpler functions, where each layer is composed of a linear graph convolutional filter nested into a nonlinearity \cite{gama2020graphs}. The graph serves as the prior about the data structure and restricts the space of functions to those exploiting this prior so to ease learning. 

GCNNs have been developed from different yet equivalent viewpoints either in the graph spectral domain or in the vertex domain. The work in \cite{bruna2013spectral} leveraged spectral graph theory to convolve the data with a learnable filter as a pointwise multiplication in the Laplacian eigenspace. Subsequently, \cite{defferrard2016convolutional, du2017topology, gama2018convolutional, xu2018powerful} built upon the shift-and-sum structure to perform convolutions directly in the vertex domain; an operation also known as finite impulse response (FIR) graph filtering \cite{shuman2013emerging, sandryhaila2013discrete}. Changing the filter type to an autoregressive moving average (ARMA) form \cite{isufi2017autoregressive}, the works in \cite{levie2017cayleynets, wijesinghe2019dfnets, isufi2020edgenets} implemented GCNNs with a rational spectral response in the convolutional layer. Differently, the work in \cite{velickovic2017graph} followed the attention idea \cite{vaswani2017attention} to aggregate nodal features, which, as it turns out, is also a FIR graph convolutional filter of order one on a graph whose edge weights are learned from data \cite{isufi2020edgenets}.

Since the graph filter is the tool that exploits the \emph{graph-data coupling} within GCNNs, most of the contributions proposed filters operating with different graph representation matrices or with different implementations. While beneficial in specific applications, this strategy focuses only on linear nodal feature aggregation, and, as such, it overshadows the implicit state-space model present in graph convolutions. Unveiling and analyzing this state-space model can bring new insight into how graph convolutions operate and allows coming up with more general nodal aggregation schemes. In fact, state-space models have been fundamental in advancing Markov chains, Kalman filtering \cite{Simon2006}, and recurrent neural networks \cite{goodfellow2016deep}. 

Inspired by the coupling between state-space models and sequential statistical learning, we put forth a similar interplay for graph convolutional filters. The state-space model considers the graph representation matrix (e.g., adjacency, Laplacian) as the state transition matrix while the intermediate nodal aggregations as system outputs (Section~\ref{sec:prelims}). We then show that the GCNN state-space convolutional module is rather limiting and propose appropriate generalizations towards a full-fledged non-linear state-space propagation rule (Section~\ref{sec:rsns}). Concretely, the contributions of this paper are twofold. First, it proposes a state-space analysis of graph convolutions, revealing that the GCNN is limited to linear nodal aggregations where the input signal of a specific layer vanishes/explodes with the filter order. Consequently, the filter coefficients have to mitigate such effect. Second, it develops a new family of graph neural networks (GNNs), which considers non-linear nodal aggregations within a layer and have intra-layer residual bridges to account for the layer input signal in the higher-order aggregations. By making parallelisms with nonlinear state-space models and with conventional RNNs, we further introduce a gating mechanism to increase the nonlinear filter order but still account for multi-resolution information in a data-driven manner. These contributions have been corroborated with numerical results in source localization and authorship attribution (Section~\ref{sec:sims}). Conclusions are drawn in Section~\ref{sec:conclusions}.


\section{Graph Convolutional neural networks}
\label{sec:prelims}



Consider a graph $\ccalG = (\ccalV, \ccalE)$ with node set $\ccalV = \{1, \ldots, N\}$, edge set $\ccalE \subseteq \ccalV \times \ccalV$, and shift operator matrix $\bbS \in \reals^{N \times N}$ such that entry $(i,j)$ satisfies $S_{ij} \neq 0$ if $(j,i) \in \ccalE$ or $i = j$. Common choices for $\bbS$ include the graph adjacency matrix $\bbA$ or the graph Laplacian matrix $\bbL$. Along with the graph, we are interested in learning from signals $\bbx = [x_1, \ldots x_N]^\top$ residing on the vertices $\ccalV$, in which entry $x_i$ corresponds to the signal at node $i$. The GSO $\bbS$ plays a role into learning from this signal because the graph encodes pairwise relationships between signal components, which in turn serves as an inductive prior.
In particular, if we consider a vector of coefficients $\bbh = [h_0, \ldots, h_{K}]^T$, we can use $\bbS$ to define graph convolutional filters as \cite{sandryhaila2013discrete}
\begin{equation}\label{eq:FIRout}
\bby = \bbh *_\bbS \bbx = \sum_{k = 0}^Kh_k\bbS^k\bbx := \bbH(\bbS)\bbx
\end{equation}
where $\bby$ is the filter output and $\bbH(\bbS) := \sum_{k = 0}^Kh_k\bbS^k$ is the filter matrix representation. 
The convolutional filter in \eqref{eq:FIRout} leverages the graph-data coupling locally. To see this, consider operation $\bbw_1 = \bbS\bbx$, which diffuses the input to neighboring vertices to produce another graph signal whose value $w_{1i}$ at node $i$ is a linear combination of signal values at the $1$-hop neighbors. Likewise, operation $\bbw_k = \bbS^k\bbx$ shifts the input $k$ times to produce a graph signal whose value $w_{ki}$ at node $i$ is a linear combination of the input signal on neighbors that are at most $k$ hops away. But since $\bbw_k$ can also be obtained from $\bbw_{k-1}$ as $\bbw_k =\bbS^k\bbx = \bbS\bbw_{k-1}$, it can be seen as the result of an aggregation from one-hop neighbor values of the former shifted signal $\bbw_{k-1}$.

Leveraging the graph convolutional filter in \eqref{eq:FIRout}, we can define graph convolutional neural networks (GCNNs) as a compositional architecture of $L$ convolutional filters and nonlinearities. At layer $l$, the GCNN takes as input a collection of $F$ signal features $\{\bbx_{l-1}^g\}_{g =1}^F$ from the previous layer, processes them in parallel with a bank of $F^2$ graph convolutional filters $\{\bbH^{fg}_l\}_{f=1}^F$, and passes these outputs through a nonlinearity to obtain the propagation rule
\begin{equation}\label{eq:convLay}
\bbx_l^f = \sigma \bigg(\sum_{g = 1}^{F}\bbH^{fg}_l(\bbS)\bbx^g_{l-1}	\bigg) = \sigma \bigg(\sum_{g = 1}^{F}\sum_{k = 0}^Kh^{fg}_{lk}\bbS^k\bbx^g_{l-1}	\bigg)
\end{equation}
for $f = 1, \ldots, F$. The $F$ outputs of layer $l$, $\{\bbx_l^f \}_f$, are inputs to the subsequent layer $l+1$, and this process repeats itself until the last layer, $l = L$, is reached. If we consider for simplicity only one input feature $\bbx_{0}:=\bbx \in \reals^{N}$ and one output feature $\bbx_L := \bbx_L^1 \in \reals^{N}$, this GCNN can be written succinctly as the map $\bbx_L = \bbPhi(\bbS; \bbx; \ccalH)$. This notation emphasizes the dependence of the parametrization on the GSO $\bbS$ and on the filter coefficients $\ccalH = \{\bbh_l^{fg}\}_{fgl}$ for all layers $l$ and filter pairs $f,g$. Graph convolutional neural networks exhibit several desirable properties. Namely, they are local and distributed information processing architectures, making them perfectly suited for distributed learning \cite{Owerko20-Power, Gama20-Distributed}. They are also permutation equivariant \cite{Gama20-Stability, ZouLerman18-Scattering} and stable to changes in the underlying graph support \cite{Gama20-Stability}. Finally, they have isomorphic properties \cite{xu2018powerful, Villar19-EquivIsomorphism, Hamilton19-Weisfeiler} and are found to be more discriminable than graph filters \cite{Pfrommer20-Discriminability}.


\subsection{The State-Space Model of Graph Convolutions}

A discrete linear system with inputs $\bbu_k \in \reals^N$ and outputs $\bby_k \in \reals^N$ can be expressed through its \textit{state-space representation} 
\begin{align}\label{eq:ssFormLinearSys}
\begin{split}
\bbw_k &= \bbA\bbw_{k-1}+\bbB\bbu_k\\
\bby_k &= \bbC\bbw_k + \bbD\bbu_k
\end{split} \qquad k = 1, \ldots, K
\end{align}
where $\bbw_k \in \reals^N$ is the system state and $\bbA$, $\bbB$, $\bbC,\bbD \in \reals^{N\times N}$ are the state-to-state, input-to-state, state-to-output, and input-to-output transition matrices, respectively.
Comparing the recursive implementation of \eqref{eq:FIRout} with \eqref{eq:ssFormLinearSys}, we see that the convolutional module of the GCNN layer can be represented as a discrete linear system where the steps $k$ correspond to graph shifts. Explicitly, the filter output $\bby = \bbH(\bbS)\bbx$ can be formulated as 
\begin{subequations}\label{eq:ssFormComp}
\begin{align}\label{eq:ssForm}
\begin{split}
\bbw_k &= \bbS\bbw_{k-1}\\
\bby_k &= h_k\bbw_k
\end{split} \qquad k = 1, \ldots, K\\
 \bby &= \sum_{k = 0}^K\bby_k \label{eq:ssFormOverall}
\end{align}
\end{subequations}
where $\bbu_k = \boldsymbol{0}$, $\bbA=\bbS$ and $\bbC=h_k\bbI$. The state is initialized as $\bbw_0 = \bbx$, and the instantaneous output as $\bby_0 = h_0\bbw_0$. The overall filter output $\bby$ is calculated as the sum of the $K+1$ instantaneous outputs $\{\bby_k\}_k$. Equation \eqref{eq:ssFormComp} makes for an interesting parallel between linear systems and graph convolutions. At the same time, it shows that the linear components of the layers of a GCNN are rather simple dynamical systems. While this is not necessarily a disadvantage, it reveals the opportunity of increasing the expressive power of GCNNs by modifying the linear system in \eqref{eq:ssFormComp}.

Moreover, we can see that if $\bbS$ has eigenvalues greater than one in magnitude, the instantaneous state $\bbw_k$ explodes. This implies that the instantaneous outputs $\bby_k$ will see little of the input signal for larger $k$. On the one hand, this will force the network coefficients $\bbh$ to learn convolutional representations of the input, and, on the other, to mitigate the explosive states for larger order states. Likewise, a similar trade-off is present if $\bbS$ has eigenvalues with magnitude less than one. In that case, we have to face with vanishing states, meaning that higher-order shifts from multi-hop neighbors will play a small role in the final output. While normalization would prevent large powers of the graph shift operator from exploding, it does not stop them from vanishing when the graph signal is aligned with eigenvectors associated with eigenvalues of magnitude less than one (i.e., any eigenvalue that is not the largest in absolute value). These trade-offs implicitly limit the degrees of freedom of the GCNN. Consequently, the model can only partially capture the coupling between the signal and the topology. This translates into limited expressive power. In the sequel, our goal is to generalize the convolutional state-space model \eqref{eq:ssFormComp} to architectures closer to a full-flagged non-linear state-space representation that still captures the coupling between the signal and the underlying topology.


\section{Nonlinear state-space extensions of GCNNs}
\label{sec:rsns} 



In this paper, we work towards extending the graph convolutional neural network layer to a propagation rule that, within a layer, can be represented as the $N$-state discrete \textit{nonlinear} system
\begin{align}\label{eq:ssRnn}
\begin{split}
\bbw_k &= \sigma_w\big(\bbA\bbw_{k-1} + \bbB\bbx_k\big)\\
\bby_k &= \sigma_y\big(\bbC\bbw_k + \bbD\bbx_k \big)
\end{split}.
\end{align}
Contrasting \eqref{eq:ssRnn} with the state space GCNN model \eqref{eq:ssFormComp}, we note three key differences. 

First, system \eqref{eq:ssFormComp} is linear in all of its components. The GCNN only applies the nonlinearity to the filter output $\bby$, but not to the shifted signals $\bbw_k$ nor to the instantaneous outputs $\bby_k$. Thus, graph convolutions limit nodal feature aggregations to the linear space. 

Second, in \eqref{eq:ssFormComp} both the state $\bbw_k$ and the instantaneous output $\bby_k$ are disconnected from the input $\bbx$. In fact, the input is only considered when initializing the state as $\bbw_0 = \bbx$. Therefore, its contribution to high-order shifts $\bbw_k$ and instantaneous outputs $\bby_k$ is small and affected by the shift operator spectra.
Additionally, nodes only learn weights to scale the influence of the values of the shifted signals $\bbS^k\bbx$ in their immediate neighborhood but leave any direct relationship with the input signal components of their $k$-hop neighbors unexploited. 

Third, while in \eqref{eq:ssFormComp} the state $\bbw_{k-1}$ is diffused through the graph to produce the next state $\bbw_k$, there is no parametric relationship between state updates; i.e., $\bbw_{k-1}$ and $\bbw_{k}$. In turn, this leads to the aforementioned instabilities related to the state-transition matrix $\bbS$. Making $\bbw_k$ a graph parametric update of $\bbw_{k-1}$ improves our control over the stability of the state-transition matrix as a whole. 

In the GNN architectures we develop next, the nodal aggregation schemes emulate a nonlinear state-space model [cf. \eqref{eq:ssRnn}] that accounts for the graph structure in a similar fashion to graph convolutions [cf. \eqref{eq:ssFormComp}]. Approaching GNNs from this state-space perspective allows changing the family of propagation rules, which are generalized from the linear form in \eqref{eq:convLay} to nonlinear node updates. As we will illustrate with the numerical experiments in Section \ref{sec:sims}, these modifications significantly improve GNN performance in a variety of application scenarios. 


\subsection{RSNs: Recursive Shift Networks} \label{sbs:rsns}

In the so-called \textit{recursive shift networks} (RSNs), we enhance the capacity of the filter \eqref{eq:ssFormComp} by making both the state $\bbw_k$ and the instantaneous output $\bby_k$ nonlinear on the state $\bbw_{k-1}$ and input $\bbx$. Explicitly, the non-linear state-space model of a RSN has the form
\begin{subequations}\label{eq:singleShift}
\begin{align}
\begin{split}
\bbw_k &= \sigma_w \bigg(h_{kww}\bbS\bbw_{k-1} + h_{kwx}\bbx	\bigg)\\
\bby_k &= \sigma_y\bigg(h_{kyw}\bbw_k + h_{kyx}\bbx\bigg)\qquad k = 1, \ldots, K
\end{split}\\
\bby &= \sigma_y\bigg(\sum_{k = 0}^K\bby_k\bigg)
\end{align}
\end{subequations}
where $h_{kww}$, $h_{kwx}$ are scalar weights encoding the dependency of state $\bbw_k$ on state $\bbw_{k-1}$ and the input $\bbx$, respectively, while $h_{kyw}$, $h_{kyx}$ are scalar weights encoding the dependency of the instantaneous output $\bby_k$ on the state $\bbw_k$ and the input $\bbx$, respectively. 
On the one hand, \eqref{eq:singleShift} retains the simplicity and efficiency of the convolutional graph filter \eqref{eq:ssFormComp}; on the other, it improves the expressive power of the graph convolution by including nonlinearities. These additional parameters as well as the nonlinearities endow the RSN with minimal degrees of freedom that are enough to control the explosion/vanishing of the state $\bbw_k$ with a better trade-off.

Despite looking similar to the conventional recurrent neural network (RNN) propagation rule, RSNs and RNNs are very different. RNNs have $N \times N$ parameter matrices, whereas in \eqref{eq:singleShift} the parameters of the nonlinear graph filters are independent of the graph dimensions. The nonlinear graph filters we consider share parameters across nodes---not shifts. This property is inherited from the graph convolutional filter [cf. \eqref{eq:ssFormComp}], in which the parameters $h_k$ are distinct for different $\bby_k$. These differences notwithstanding, we leverage the analogy with RNNs to consider gating mechanisms in Section \ref{sbs:lssms}. 


\subsection{LSSMs: Long Short Shift Memories} \label{sbs:lssms}

In both \eqref{eq:ssFormComp} and \eqref{eq:singleShift}, the filter order $K$ controls the information locality in the vertex and spectral domains. In the vertex domain, the order implies that state $[\bbw_K]_i$ at node $i$ receives information from nodes up to $K$ hops away; i.e., it defines a ``local window'' around the nodes. In the spectral domain, it controls the sharpness of the filter frequency response \cite{shuman2013emerging, sandryhaila2013discrete}. When the information at a particular layer is localized around a few graph frequencies (eigenvalues of $\bbS$), higher filter orders are needed; i.e., the filter order imposes a ``local window'' around the graph frequencies. This is in agreement with the uncertainty principle of signal localization \cite{agaskar2013spectral, tsitsvero2016signals, teke2017uncertainty}, which states that low values of $K$ correspond to localized windows in the vertex domain, but not in the spectral domain (and vice-versa).

Increasing the filter order is thus necessary to capture more information in the vertex domain while retaining localized responses in the spectral domain. However, large $K$ usually leads to numerical instabilities associated with large powers $\bbS^K$ and, depending on the eigenvalues of $\bbS$, we also have to cope with vanishing or exploding gradients. These challenges are similar to those encountered in RNNs. There, they are typically addressed by gating mechanisms that introduce an additional set of parameters to control how the information propagates in different state updates \cite{hochreiter1997long}. Here, we will use gates \emph{within} the GNN layer to capture long-range dependencies over the graph because of the high order of \eqref{eq:singleShift}. 

In analogy with long-short term memories (LSTMs), we call our architecture \emph{long-short shift memory} (LSSM). LSSMs comprise learnable gating parameters taking values in the interval $[0,1]$. These parameters control the information passed to state $\bbw_k$ and instantaneous output $\bby_k$ in \eqref{eq:singleShift}. An LSSM filter comprises:
\begin{itemize}
\item Updating the $N\times 1$ internal memory variable $\tbc[k]$ as
\begin{equation}\label{eq:insMem}
\tbc_k = \tanh\big(	h_{kcw}\bbS\bbw_{k-1} + h_{kcx}\bbx	\big)
\end{equation}
to track the state update.
\item Updating the $N \times 1$ forget gate $\bbgamma_\text{f}[k]$, update gate $\bbgamma_\text{u}[k]$, and state gate $\bbgamma_\text{w}[k]$ respectively as
\begin{subequations}\label{eq:gatesLSSM}
\begin{align}
\bbgamma_{\text{fk}} &= \text{sigmoid} \big(h_{kfw}\bbS\bbw_{k-1} + h_{kfx}\bbx	\big)\\
\bbgamma_{\text{u}k} &= \text{sigmoid} \big(h_{kuw}\bbS\bbw_{k-1} + h_{kux}\bbx	\big)\\
\bbgamma_{\text{w}k} &= \text{sigmoid} \big(h_{kww}\bbS\bbw_{k-1} + h_{kwx}\bbx	\big)
\end{align}
\end{subequations}
which are internal variables that track the system evolution with their own set of parameters. The sigmoid nonlinearity ensures that the values are in the interval $[0,1]$.
\item Updating the $N\!\times\!1$ global memory variable $\bbc_k$ and state $\bbw_k$
\begin{subequations}\label{eq:memStateLSSM}
\begin{align}
\bbc_k &= \bbgamma_{\text{f}k}~\circ~\bbc_{k-1} + \bbgamma_{\text{u}k}~\circ~\tbc_k \\
\bbw_k &= \bbgamma_{\text{w}k} \circ \tanh(\bbc_k)
\end{align}
\end{subequations}
where $``\circ"$ denotes the element-wise Hadamard product. The forget gate $\bbgamma_{\text{f}k}$ and update gate $\bbgamma_{\text{u}k}$ control which entries of the former global memory $\bbc_{k-1}$ to propagate and which entries to update through the internal memory $\tbc_k$ [cf. \eqref{eq:insMem}]. The global memory variable is then used to update the state $\bbw_k $, whose value is in turn controlled by the state update gate $ \bbgamma_{\text{w}k}$.
\item Setting the instantaneous output $\bby_k$ to
\begin{equation}\label{eq.LSSS_instOut}
\bby_k = \sigma_y\bigg(h_{kyw}\bbw_k + h_{kyx}\bbx	\bigg).
\end{equation}
\item Setting the overall LSSM output to
\begin{equation}\label{eq:outLSSM}
\bby = \sigma_y\big(\sum_{k = 0}^K\bby_k\big).
\end{equation}
\end{itemize}

In summary, the LSSM filter is defined by steps \eqref{eq:insMem}--\eqref{eq:outLSSM}. Note that this \emph{nonlinear graph filter} updates the state $\bbw_k$ as a nonlinear, shifted version of the former state while prioritizing information coming from certain nodes and, thus, only learning state updates on nodes that are relevant for the task at hand. The update is controlled by the gating mechanisms [cf. \eqref{eq:gatesLSSM}], which are graph-based state-space models themselves. The additional parameters further increase the LSSM degrees of freedom compared with RSNs, allowing to control the state updates and also to learn where a higher vertex-spectra locality is needed. Substituting $\bbH^{fg}_l(\bbS)$ for the LSSM filter in \eqref{eq:convLay} leads to an LSSM-GNN layer update rule. Because of gating, the LSSM-GNN can be parametrized with higher values of $K$. This \emph{longer memory} over the graph endows the LSSM-GNN with a better accuracy-robustness trade-off than the GCNN and the RSN.

\begin{remark}
Regarding the number of parameters of the proposed state-space extensions, they indeed have more parameters per layer than the GCNN, but the difference is not expressive --- the RSN has $4KF^2$ [cf. equation \eqref{eq:singleShift}] and the LSSM $10(K+1)F^2$ parameters per layer [cf. equations \eqref{eq:insMem}--\eqref{eq:outLSSM}], versus $KF^2$ for the GCNN. 
\end{remark}


\section{Numerical Experiments}
\label{sec:sims}



In the following, we describe the scenarios and respective experimental setups used to corroborate the proposed solutions. The baseline setups are those of \cite{isufi2020edgenets}, which compares the GCNN with different state-of-the-art approaches. The models we evaluate are: $(i)$ the conventional GCNN with linear filters [cf. \eqref{eq:ssFormComp}]; $(ii)$ the RSN [cf. \eqref{eq:singleShift}]; $(iii)$ the LSSM [cf. \eqref{eq:insMem}-\eqref{eq:outLSSM}]. All models have ReLU nonlinearities between layers and are trained using the ADAM optimizer with parameters $\beta_1 = 0.9$ and $\beta_2 = 0.999$ \cite{kingma2014adam}.

\smallskip
\noindent\textbf{Source localization.} The goal of this experiment is to identify the source community of a signal diffused over the graph given a snapshot of the signal at an arbitrary time step. The graph is an undirected stochastic block model (SBM) with $N = 50$ nodes divided into $C = 5$ blocks, each representing a community $\{c_1, \ldots, c_5\}$. The intra-community probability is $p = 0.8$ and the inter-community probability is $q = 0.2$. The source signal $\bbx[0]$ is a Kronecker delta centered at one source node and diffused at time $t \in [0, 50]$ as $\bbx[t] = \bbS^t\bbx[0]$, where $\bbS$ is the graph adjacency matrix normalized by the maximum eigenvalue. The training set comprises $10,240$ tuples of the form $(\bbx[i], c_i)$ for a random $t$ and $i \in \{1, 2, 3, 4, 5\}$. The validation and the test sets are both composed of $2,560$ tuples. The models are trained with batch sizes of $100$ samples for $40$ epochs and a learning rate $10^{-3}$, which is tuned for the GNN but not for the proposed models. The performances are averaged over ten different graph realizations and ten data splits, for a total of $100$ realizations.

We vary the filter order $K$ in the set $\{4, 16, 32\}$ to compare the accuracy-robustness trade-off of the GCNN, RSN, and LSSM in the source localization scenario. All architectures have $L=1$ layer and $F=4$ features. The nonlinearities $\sigma_w$ and $\sigma_y$ are the ReLU in \eqref{eq:singleShift}; and $\sigma_y$ is the ReLU in \eqref{eq.LSSS_instOut} and \eqref{eq:outLSSM}. At the output of each architecture, a readout layer maps the output signal to a one-hot vector of dimension $C$, which is then fed to a softmax.

\begin{figure}[t]
\centering
    \includegraphics[width=\columnwidth]{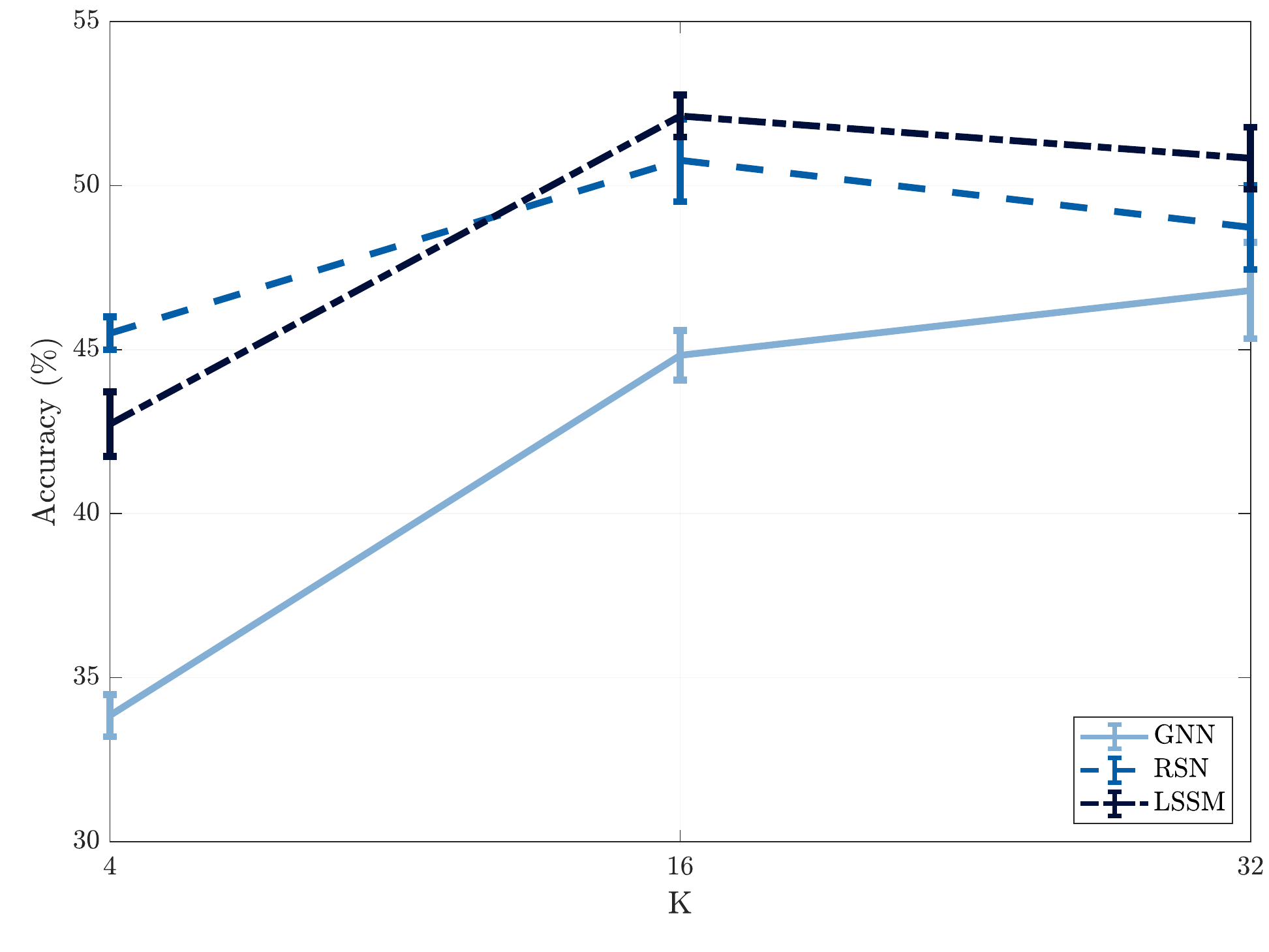}
    \caption{Source localization accuracy for $K=4$, $16$ and $32$. Error bars have been scaled by $0.5$.}
    \label{fig:source_loc}
\end{figure}

The average test accuracies are shown in Figure \ref{fig:source_loc}. Both the RSN and the LSSM outperform the GCNN by a significant margin for all values of $K$. While the RSN achieves the best accuracy when $K$ is fixed at $K=4$, the LSSM exhibits a better performance for larger values of $K$, which indicates its better robustness-accuracy trade-off for high-order filters. This is further validated by the fact that they present the smallest standard deviation for $K=16$ and $K=32$.

In a second experiment, we aimed at verifying if increasing the number of layers of the GCNN (which results in adding more nonlinearities to the architecture) would yield similar gains in performance to those observed for the RSN and LSSM in Figure \ref{fig:source_loc}. We set $K=4$ and $F=4$ and train all architectures for $L=2$ and $L=4$. Average results are shown in Table \ref{tab:sourceloc} for 5 graph realizations and 5 dataset realizations each. Once again, we observe that the LSSM achieves the best accuracy and that both state-space extensions outperform the GCNN.

%

\begin{table}[t]
    \centering
    \caption{Source localization accuracy for multiple layers (\%).}
    \label{tab:sourceloc}
    \begin{tabular}{l|cc}
        \hline
        		& $L=2$    			& $L=4$         	\\ \hline
        GCNN & $49  \pm 5$ 		& $40  \pm 17$  	\\
        RSN  & $62  \pm 5$ 		& $53  \pm 12$  	\\
        LSSM & $\mathbf{62 \pm 3}$ 		& $\mathbf{60  \pm 16}$ 	\\ \hline
    \end{tabular}
\end{table} 

\begin{figure}[t]
    \centering
    \includegraphics[height=0.31\textheight]{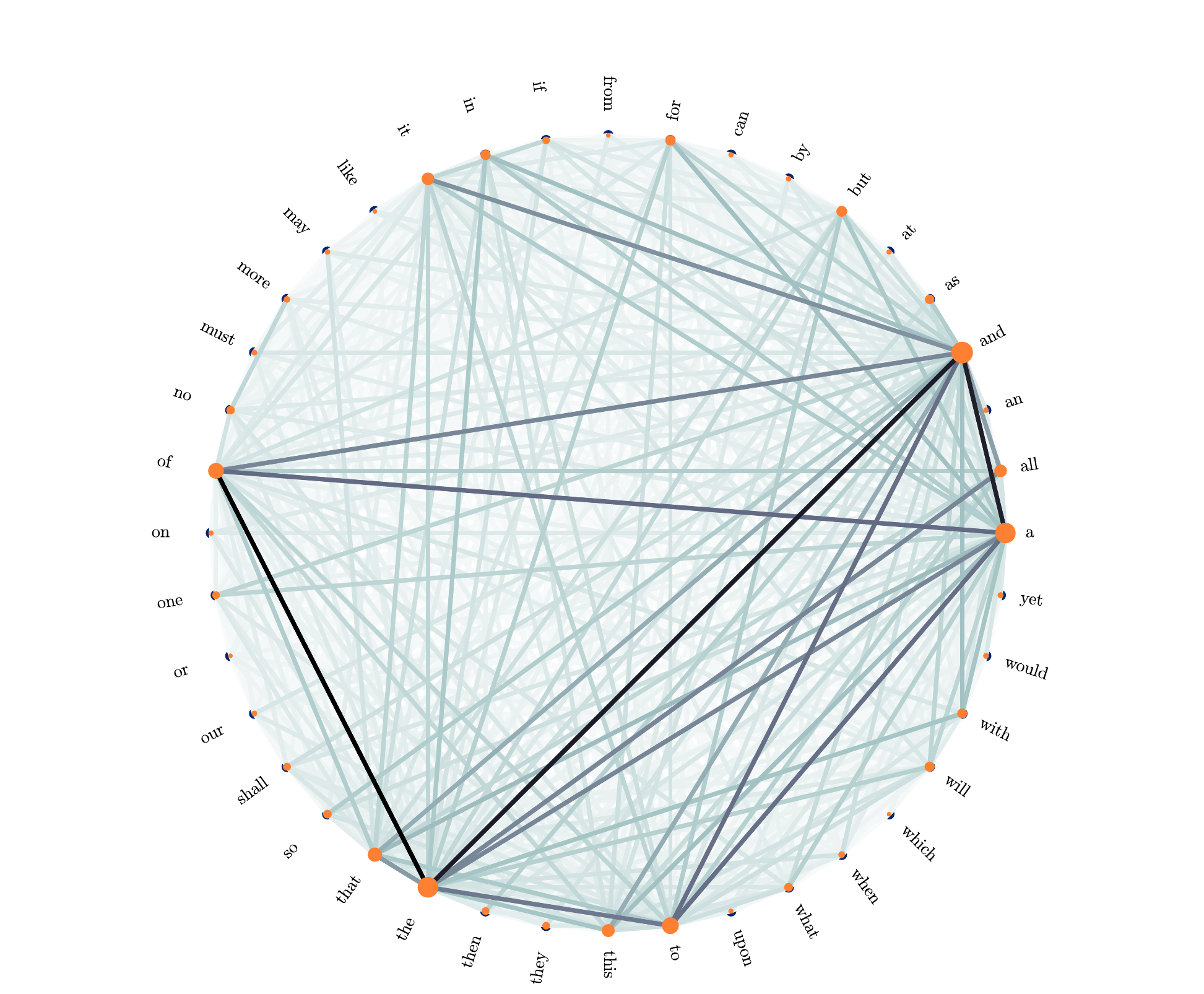}
    \caption{Example of word adjacency network for the author Robert Louis Stevenson.}
    \label{fig:wan}
\end{figure}

\noindent\textbf{Authorship attribution.} In authorship attribution, the learning task is to decide whether a $1,000$-word text excerpt has been authored by a specific author or by any of the other $20$ contemporary authors in the author pool, given their word adjacency network (WAN) \cite{segarra2015authorship}. WANs are author-specific directed graphs whose nodes are function words without semantic meaning (e.g., prepositions, pronouns, conjunctions) and whose directed edges capture the transition probabilities between pairs of function words. An example of WAN is shown in Figure \ref{fig:wan} and the graph signal is the word frequency count.

Like in \cite{isufi2020edgenets}, we classify texts for: Jane Austen, Emily Bronte, and Edgar Alan Poe. The WANs of these authors have from $N = 190$ to $N = 210$ function word nodes. We consider a train-test split of $95\% - 5\%$ of the available texts per author and around $8.7\%$ of the train samples are used for validation. This leads to around $1,000$ training samples and $100$ validation and test samples. For each author, we extend the training, validation, and test sets by the same number of text samples taken at random from the author pool. All models are trained with batches of $100$ samples for $25$ epochs, and the learning rate is $5 \times 10^{-3}$. The loss function is the cross-entropy and we report average test accuracies for $30$ data splits.

In this experiment, we fix the parameters to those of the $1$-layer GCNN which achieved the best performance in the source localization experiment---$F=4$ and $K=32$---to make for a fair comparison. The results are presented in Table \ref{tab:authorship}. We observe that the RSN outperforms the GCNN for all authors except Br\"onte, and that the LSSM exhibits the best performance by a large margin.

%


\begin{table}[t]
    \centering
    \caption{Authorship attribution accuracy (\%).}
    \label{tab:authorship}
    \begin{tabular}{l|ccc}
        \hline
        & Austen        & Bronte        & Poe          \\ \hline
        GCNN  & $64  \pm 20$ & $71  \pm 16$ & $64 \pm 18$ \\
        RSN  & $80  \pm 19$ & $68  \pm 16$ & $72 \pm 19$ \\
        LSSM & $\mathbf{87  \pm 14}$ & $\mathbf{73  \pm 15}$ & $\mathbf{83 \pm 14}$ \\ \hline
    \end{tabular}
\end{table}


\section{Conclusions}
\label{sec:conclusions}



Implicitly, graph convolutional neural networks carry a state-space model in their convolutional update rule. In this paper, we make this relationship explicit and analyze its behavior from an internal state perspective. By noting that the internal state may explode or vanish depending on the spectrum of the shift operator, we argued that this places a burden on the GCNN parameters because they need to be learned to control such phenomena, leading to a stability-performance trade-off. We then built further links with discrete state-space models to develop a new family of graph neural networks, in which nodal aggregations are performed in a nonlinear and parametric manner. The latter leads to higher degrees of freedom to control the stability-performance trade-off and allows developing new solutions to improve the expressive power of GCNNs. We proposed two such solutions, namely, $i)$ a recursive shift network that includes the input signal in every state update; $ii)$ a long-short term shift memory that allows further increasing the filter order within a layer through the introduction of gating mechanisms akin to the gates in conventional LSTMs. Numerical results on source localization and authorship attribution validate both models. In the future, we plan on investigating the theoretical benefits of these nonlinear aggregation rules.


\bibliographystyle{bibFiles/IEEEbib}
\bibliography{bibFiles/myIEEEabrv,bibFiles/biblioRSN}

\end{document}